\documentclass[aps,prb,showpacs,twocolumn]{revtex4}
\usepackage{graphicx,psfrag}
\usepackage{amssymb}
\usepackage{amsmath}
\usepackage{pstricks}
\usepackage{pst-plot}
\usepackage{pst-grad}

\newcommand{\be}{\begin{equation}}  \newcommand{\ee}{\end{equation}}
\newcommand{\bea}{\jot1ex\begin{eqnarray}}  \newcommand{\eea}{\end{eqnarray}}
\newcommand{\NL}{\nonumber\\}
\newcommand{\mm}[2]{{\renewcommand{\arraycolsep}{0.2em}%
             \left(\begin{array}{#1}#2\end{array}\right)}}
 \newcommand{\vK}{{\mathbf K}}
\newcommand{\vx}{{\mathbf x}} \newcommand{\vvr}{\mathbf r}
\newcommand{\mA}{{\cal A}} \newcommand{\mB}{{\cal B}} 
\newcommand{\mD}{{\cal D}} \newcommand{\mS}{{\cal S}}
\newcommand{\eq}[1]{Eq.~(\ref{#1})}

\begin{document}
\title{Electric transport through circular graphene quantum dots:
  Presence of disorder}
\author{G. Pal}
\author{W. Apel}
\author{L. Schweitzer}
\affiliation{Physikalisch-Technische Bundesanstalt (PTB),
Bundesallee 100, 38116 Braunschweig, Germany}
\begin{abstract}
The electronic states of an electrostatically confined cylindrical
graphene quantum dot and the electric transport through this device 
are studied theoretically within the continuum Dirac-equation
approximation and 
compared with numerical results obtained from a tight-binding lattice
description. A spectral gap, which may originate from strain effects, 
additional adsorbed atoms or substrate-induced sublattice-symmetry 
breaking, allows for bound and scattering states. As long as the 
diameter of the dot is much larger than the lattice constant, the
results of the continuum and the lattice model are in very good
agreement. We also investigate the influence of a sloping
dot-potential step, of on-site disorder along the sample edges, of
uncorrelated short-range disorder potentials in the bulk, and of random
magnetic-fluxes that mimic ripple-disorder. The quantum dot's spectral
and transport properties depend  
crucially on the specific type of disorder. In general, the peaks in
the density of bound states are broadened but remain sharp only in the
case of edge disorder.  
\end{abstract}
\date{\today} 
\maketitle

\section{Introduction}
Recent advances in the fabrication of single layer graphene structures
have facilitated the realization and application of graphene
nanoelectronic devices. Graphene nanoribbons (GNR)\cite{HOZK07} with
constant or varying widths, down to sub-10\,nm, have already been
prepared and operated as field effect transistors.\cite{LWZLD08}  
Also, quantum dots (QD), which were either plasma etched or carved out
mechanically have been investigated
recently.\cite{Pea08,Sea08,TCAG09,Gea09,LRGSB10}  
The usual way to confine QDs electrostatically by external gates, 
however, is in general not practicable in pristine graphene because of
the gap-less bandstructure. Also, the charge carriers behave like
Dirac-fermions and thus the occurrence of true bound states is
affected by the Klein-tunneling mechanism.\cite{KNG06,PPCF10} The
latter can be seen experimentally by measuring the current flowing
through a sample with a tunable potential barrier across the graphene
sheet.\cite{HSSTYG07,SHG09,YK09,VLBL09} 

Therefore, physical or chemical effects that are able to open a gap in
the energy spectrum of single layer graphene are of vital
importance for further electric applications. Among the proposed
mechanisms for the creation of a spectral gap, are size quantization in
armchair GNR, magnetic interactions\cite{SCL06a} between the edge
states, as well as application of external electric potentials along
the sample edges\cite{APS11} in zigzag GNR. Other proposed mechanisms 
that are effective also in broad graphene sheets are strain-induced
gap openings,\cite{GLZ08,PCP09} substrate-induced band gap
formation,\cite{GKBKB07} and chemical effects of adsorbent atoms and 
molecules.\cite{RPCB08} A gap opening attributed to a breaking of the
sub-lattice symmetry and detected in epitaxial graphene grown on a SiC
substrate was shown to produce a gap of about 0.26\,eV.\cite{Zea07}     

In the absence of a spectral gap it was theoretically shown that an
electrostatically confined QD can accommodate only quasi-bound 
states.\cite{SE07,MP08, HA08}  
At the Dirac point, i.e., at energy $E=0$ where valence and conduction
band touch, the electronic transport through QDs of certain shapes
was also considered.\cite{BTB09} In this special situation, sharp
resonances in the two-terminal conductance were predicted.
However, in the presence of a spectral gap around the Dirac point, 
which in principle can be created by one of the mechanisms mentioned
above, true bound states were obtained.\cite{TBLB07,RNBT09}  
Additional energy gaps due to Landau quantization can also be induced
in quantum dot physics by the application of a strong perpendicular
magnetic field.\cite{SESI08,RNBT09} 

In this paper, we investigate the electric transport through a
circular electrostatic potential in the presence of a spectral gap.
In mesoscopic physics, a positive scattering potential is usually
called a quantum anti-dot, and a quantum dot when the potential is
negative. Owing to the chiral symmetry of graphene, our theory is
valid in both cases, and for the sake of simplicity we always call 
it a quantum dot. We calculate analytically the transport
cross-section derived from the continuum Dirac formulation and 
compare the results with the two-terminal conductance obtained
numerically from a tight-binding (TB) lattice model. 
In doing so, we also study the limit of validity of the
continuum Dirac fermion description that is considered to be a good
approximation of the low energy physics in the vicinity of the Dirac
point. 

Real graphene samples are hardly perfect due to the presence of  
disorder induced by adsorbent\cite{WSSLM08,Bea09} atoms and molecules,
which mainly affect the unsaturated dangling bonds at the sample
edges, or due to bulk defects and ripples.\cite{Mea06,MG06,Mea07,GKV08} 
Therefore, we investigate the influence of such disorder effects on
the spectral and transport properties of an electrostatically confined
graphene quantum dot. In particular, we find that the impact of
one-dimen\-sional uncorrelated random disorder potentials, which only 
disturb the a-sublattice sites at one edge and the b-sublattice sites 
at the
other, causes changes of the quantum dot properties that are different
from the case of short-range or random magnetic flux bulk disorder.
In addition, we analyze what happens to the bound states' energies when 
the boundary of the electrostatic potential confining the QD varies
over a length of several lattice constants. The effect of disorder on
the single-particle states at the edges of graphene QDs has already
been discussed recently.\cite{WAG10} 

The outline of the paper is as follows. In section \ref{SDM}, we
specify the continuum Dirac model and the TB lattice model for
graphene, both include an electrostatic potential defining the QD.  
For a piecewise constant radially symmetric potential, the Dirac
equation is analytically 
tractable. We discuss the character of the electronic eigenstates
occurring in various regions of an energy versus potential diagram. 
In section \ref{SBS}, we obtain the energies of the bound states of
the isolated QD and compare them with numerical results of the
corresponding TB lattice model. In section \ref{SET}, we add an
environment to the isolated QD and calculate both the scattering
cross-section and, more specifically, the transport cross section and 
compare it to numerical calculations of the two-terminal conductance
obtained with a transfer matrix method. Finally, section \ref{disorder}
is devoted to a numerical study of various types of disorder and its
implications on the density of states in the vicinity of the Dirac point.

\section{Graphene quantum dot with radially symmetric potential \label{SDM}}
\subsection{Continuum model}
In the following, we recall the basic notions of the Dirac description of
graphene. It stems from the low-energy expansion of the
TB Hamiltonian around the $\vK$ and $\vK'$ points in the first
Brillouin zone where the conduction and valence band touch.
The wavefunction is then a four-dimensional spinor
$\mm{cccc}{\xi_{+}&\xi_{-}&\eta_{+}&\eta_{-}}$, $\xi$ denotes the
wavefunction in the valley $\vK$, and $\eta$ that in the valley
$\vK'=-\vK$ and the indices $+,-$ denote the sublattice (the unit cell
contains two points). There is no coupling between the valleys and
thus we reduce the Dirac equation to a $2\!\times\!2$ matrix form for
the valley $\vK$ 
\be
\mm{cc}{
E-\Delta-V(r) & i\partial_x - \partial_y \\ 
i\partial_x + \partial_y & E+\Delta-V(r) }
\mm{c}{\xi_{+}(\vx) \\ \xi_{-}(\vx) } =0.  \label{SG}
\ee
The states $\eta_{\pm}$ in the other valley $\vK'$ are obtained by the 
transformation $x\to -x$ and they follow a  similar treatment as
$\xi_{\pm}$. In \eq{SG}, $E$ is the energy and $V(r)$ the radially
symmetric potential. Also, we introduced a constant mass term
$\Delta$ that will account for a gap $2\Delta$ in the energy
spectrum. The gap is assumed to be substrate-induced as found in
epitaxial graphene.\cite{Zea07}  
Above we have set $\hbar v_{\rm F}=1$, however, in physical units, the
Fermi velocity $v_{\rm F}=3ta/(2\hbar)$ is obtained from the TB model
where $a=1.42\,$\AA{} is the carbon-carbon distance, and
$t\approx 2.7$\,eV is the hopping integral. 

\begin{table}
\caption{\label{tabeqs}Choice of the Bessel functions $I_{m}(x)$,
  $J_{m}(x)$, $K_{m}(x)$, $N_{m}(x)$ with
  $H^{(1,2)}_{m}(x)=J_{m}(x)\pm iN_m(x)$ that describe the radial part
  of the electronic wavefunctions inside and outside the quantum dot.} 
\begin{ruledtabular}
\begin{tabular}{c|c|c}
  & inside $r<R$ & outside $R<r$ \\ \hline
I & $k'=i\kappa'$ & $k=i\kappa$ \\
  & $\kappa'=\sqrt{\Delta^2-(E-U)^2}$ &  $\kappa=\sqrt{\Delta^2-E^2}$ \\
  & $\xi_{J;\pm}(r) \propto I_{J\pm 1/2}(\kappa'r)$ & $\xi_{J;\pm}(r) 
    \propto K_{J\pm 1/2}(\kappa r)$ \\ \hline
II & $k'=\sqrt{(E-U)^2-\Delta^2}$ &  as in I \\
   & $\xi_{J;\pm}(r) \propto J_{J\pm 1/2}(k'r)$ & \\ \hline
III & as in I &  $k=\sqrt{E^2-\Delta^2}$ \\ 
    &  &  $\xi_{J;\pm}(r) \propto H^{(1,2)}_{J\pm 1/2}(k r)$ \\ \hline
IV  & as in II & as in III
\end{tabular}
\end{ruledtabular}
\end{table}

Since \eq{SG} is rotationally invariant, we decompose the wavefunction
$\xi$ into components with fixed angular momentum 
\be
 \xi_{\pm}(\vx) = \sum_J e^{i\varphi (J\pm\frac{1}{2})} \; \xi_{J;\pm}(r) 
\ee 
where $J$ is  half-integer.
Then, for a piecewise constant $V(r)$, the remaining radial equation
is easily solved in terms of the Bessel functions $I_{m}(x)$,
$J_{m}(x)$, $K_{m}(x)$, $N_{m}(x)$, and
$H^{(1,2)}_{m}(x)=J_{m}(x)\pm iN_m(x)$. We consider here $V(r) = U
\Theta(R-r)$, where $U$ denotes the potential in the QD and $R$ its
radius (measured in units of $a$).  
The Bessel functions that describe the electronic states inside and
outside the dot are chosen according to whether the local energy is
outside or inside the gap, i.e., if the wavevectors $k'$ (for $r<R$)
and $k$ (for $r>R$) are purely real or imaginary, respectively, as
summarized Table~\ref{tabeqs}.

\begin{figure}[b]
\includegraphics[width=0.75\columnwidth]{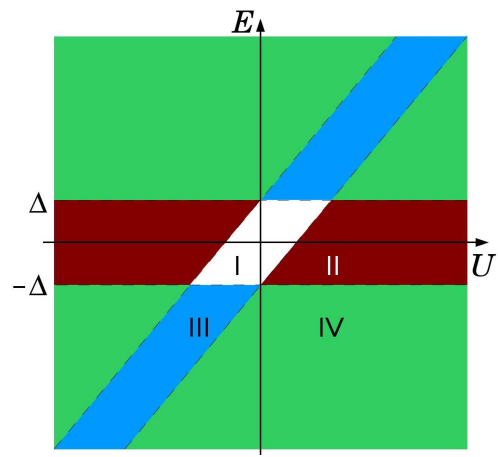}
\caption{(Color online) Energy domains for the QD states. I (white): no
  states. II (black, red): bound states. III (gray, blue):
  tunneling states. IV (light gray, green): scattering states.} \label{QDZ} 
\end{figure}

In the 'energy vs.\ potential' diagram, as shown in Fig.~\ref{QDZ}, one
distinguishes between four different behaviors of the wavefunction.
The diagram is symmetric under the transformations $E\to -E$ and $U\to
-U$, and can be understood as follows. Everywhere in the diagram, the
wavevectors inside and outside the dot satisfy $k'^2=(E-U)^2-\Delta^2$
and $k^2=E^2-\Delta^2$, respectively. In domain I, one has
$|E-U|<\Delta$ and $|E|<\Delta$, and consequently $k'^2<0$ and
$k^2<0$. Therefore, both wavevectors $k'$ and $k$ are purely imaginary
and the wavefunction decays inside and outside the  QD. 
The  domain II consists of all the $E$ and $U$ values that satisfy the
$|E-U|>\Delta$ and  $|E|<\Delta$ inequalities ($k'^2>0$ and $k^2<0$), 
so that $k'$ is real and $k$ is purely imaginary. True bound states
that oscillate for $r<R$ and decay as $\exp{(-r\sqrt{\Delta^2-E^2})}$ 
when $r\rightarrow\infty$, can exist therefore only in the energy domain
II. 
For domain III one has $|E-U|<\Delta$ and  $|E|>\Delta$, i.e.,
$k'^2<0$ and $k^2>0$. Here, the wavevector $k'$ is imaginary and $k$
is real, so that the wavefunction decays inside and oscillates outside
the QD (tunneling regime). 
The domain IV is given by $|E-U|>\Delta$ and  $|E|>\Delta$, i.e.,
$k'^2>0$ and $k^2>0$. Therefore,  both $k'$ and $k$ are real, and
the wavefunction oscillates inside and outside the QD (scattering regime).

In the domain I, the Bessel functions that satisfy the Dirac equation
are ${\cal A'}_{\pm} I_{J\pm 1/2}(\kappa' r)$ for $r<R$ and ${\cal
  A}_{\pm} K_{J\pm 1/2}(\kappa r)$ for $r>R$. Here, ${\cal A'}_{\pm}$
and ${\cal A}_{\pm}$ are the amplitudes of the wavefunction and
$\kappa$ and $\kappa'$ are the wavevectors inside and outside the QD,
respectively. 
The other two Bessel functions are excluded as possible solutions
because they diverge for $r=0$ and $r\rightarrow \infty$, leading to
infinite densities at the origin and non-normalizable wavefunctions. 
The amplitudes are determined from the Dirac equation and from the
matching conditions at $r=R$ for the wavefunctions inside and outside
the QD. The matching condition leads to an equation for ${\cal
  A'}_{\pm}$ and ${\cal A}_{\pm}$ that has no nonzero real solutions 
for $\kappa$ and $\kappa'$,
and hence, there can be no states that decay inside and outside the
QD. In the following sections, we discuss the domains II--IV.

\subsection{Lattice model}
The single band TB Hamiltonian that describes the non-interacting
electrons in graphene in the presence of a radially symmetric
potential reads 
\be
\hat H_{\text{TB}}=\sum_{\vvr} (U_{\vvr}+\Delta_{\vvr})
c_{\vvr}^{\dag} c_{\vvr}^{} -
t\sum_{\left< \vvr\ne \vvr' \right>} c_{\vvr}^{\dag} c_{\vvr'}^{}.
\label{TBham}
\ee
Here, $c_{\vvr}^{\dag}$ and $c_{\vvr}$ are the fermionic particle
creation and annihilation operators at the sites $\vvr$ of a hexagonal
lattice with carbon-carbon distance $a$. The $\left< \vvr\ne \vvr'
\right>$ denote nearest-neighbor sites and $t=2.7$\,eV is the hopping
parameter. In the following, we set the energy scale by putting $t=1$. 
The potential is $U_{\vvr}=U$ for sites inside the QD ($r<R$)
and $U_{\vvr}=0$ outside. Also, we have introduced a lattice anisotropy
$\Delta_{\vvr}$ for the two sublattices of graphene, $\Delta_{\pm}=\pm
\Delta$. This anisotropy opens an energy gap $2\Delta$ in the spectrum
and allows for the existence of true bound states. In the presence 
of disorder to be considered later, the above Hamiltonian is modified
to account also for random on-site potentials or for random transfer
terms between nearest neighbor sites.

\section{Bound states of the isolated QD \label{SBS}}
\subsection{Continuum model}
\begin{figure}
\includegraphics[width=1.\columnwidth]{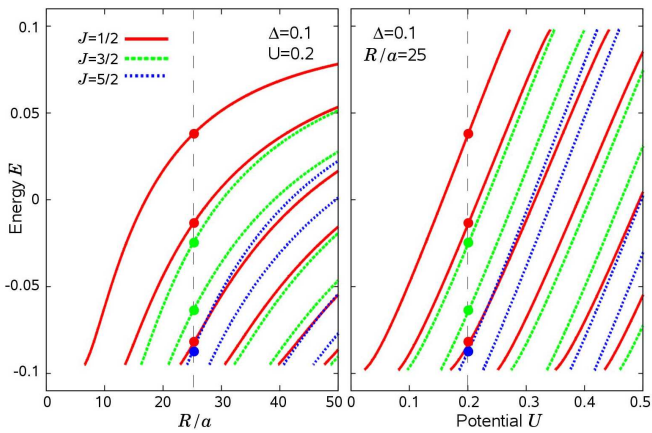}
\caption{(Color online) The energies of the bound states in a graphene
  QD as function of the dot radius (left panel) and potential strength
  (right panel). We use  $\Delta=0.1$ and show the results for $J=1/2$
  (solid line), $J=3/2$ (dashed line), and $J=5/2$ (dotted line)
  only. The intersections of the solutions with the vertical lines
  yield the  bound states energies for $R=25\, a$ and $U=0.2$. 
\label{BoundStates1}}
\end{figure}

In this section, the energies of the bound states are calculated.
To this end, we put $V(r)=U \Theta(R-r)$ and consider the region II
in Fig.~\ref{QDZ}.  The solution of the radial Dirac equation for
$r \gtrless R$ is given in terms of Bessel functions as
\be
 \xi_{\pm}(r) =
\left\{ \begin{array}{ll}
{\mA}'_{\pm}\,J_{J\pm\frac{1}{2}}(k' r), & r<R  \\
{\mA}_{\pm}\,K_{J\pm\frac{1}{2}}(\kappa r), & r>R .
 \end{array} \right.
\ee
The other two Bessel functions are excluded as solutions: for $r\to 0$,
$N_{J\pm\frac{1}{2}}(k' r)$ diverges and the density would be infinite
at the origin, for large $r$, $I_{J\pm\frac{1}{2}}(\kappa r)$
diverges and the wavefunction would not be normalizable.
The amplitudes ${\mA}_{-}$ and ${\mA}'_{-}$ are tied to ${\mA}_{+}$ and
${\mA}'_{+}$, respectively, by the equations of motion
\begin{equation}
\begin{array}{rl}
\left( \begin{array}{cc}  E-U-\Delta & -ik' \\ ik' & E-U+\Delta \end{array} \right)
\left( \begin{array}{c}  {\mA}'_{+} \\ {\mA}'_{-} \end{array} \right) =0, & r<R \\
\left( \begin{array}{cc}  E-\Delta & -i\kappa \\ -i\kappa & E+\Delta \end{array} \right) 
\left( \begin{array}{c}  {\mA}_{+} \\ {\mA}_{-} \end{array} \right)=0, & r>R.
\end{array}
\end{equation}
There are two matching conditions at $r=R$. Thus one obtains the
following condition for the existence of a non-zero wavefunction
(bound state) 
\be
 \sqrt{\frac{E-U+\Delta}{E-U-\Delta}}\;
 \frac{J_{J+\frac{1}{2}}(k' R)}{J_{J-\frac{1}{2}}(k' R)} =
 - \sqrt{\frac{\Delta+E}{\Delta-E}}\;
 \frac{K_{J+\frac{1}{2}}(\kappa R)}{K_{J-\frac{1}{2}}(\kappa R)}.
 \label{BS}
\ee
The other valley yields a corresponding equation with
$(J+1/2)$ replaced by $-(J-1/2)$.
Since we have to consider all half-integer $J$'s in \eq{BS}, each 
bound state is doubly degenerate when one considers both valleys.
In Fig.~\ref{BoundStates1}, we show the evaluation of the energies 
of the bound states vs.\ the QD radius and vs.\ the value of  
the potential. Since everything is symmetric under $E\to -E$ and
$U\to -U$ as mentioned above, we show only states for $U\geq 0$.
For clarity, we present only the states with $J=1/2$,  $3/2$, and
$5/2$. Here, we use $\Delta=0.1$ in TB units $t=1$ (which corresponds
to $\Delta=0.27$\,eV). Note that when $R=25\, a$  and $U=0.2$, there
are six bound states in the QD: three with $J=1/2$, two with $J=3/2$,
and only one with $J=5/2$, as shown in Fig.~\ref{BoundStates1}. 
When the QD radius or the strength of the confinement potential
increases, more bound states can be accommodated in the QD.
Our results are in excellent agreement  with previous calculations of 
bound state energies in a radially-symmetric\cite{RNBT09} or a
rectangular-shaped\cite{TBLB07} graphene QD.

\subsection{Lattice model}
It is well known that the above results obtained within the commonly
applied continuum Dirac-fermion approximation have a limited range of
validity. To investigate this in more detail and to check
possible deviations originating from neglecting lattice effects,
we compute the eigenvalues of the lattice model via a numerical
diagonalization of the TB Hamiltonian (\ref{TBham}). In our
calculations, we study a graphene ribbon consisting of $2 N_{\rm ac}
\times N_{\rm zz} =2\cdot 100 \times 100$ carbon atoms with 
periodic boundary conditions ($N_{\rm ac}$ and $N_{\rm zz}$ are the
number of armchair and zigzag lines, respectively). For the lattice 
anisotropy, we take $\Delta_{\pm}=\pm 0.1$ and calculate
the eigenvalues for several QD radii as a function of the potential
strength. Changing $N_{\rm ac}$ or $N_{\rm zz}$ does not alter the
resulting eigenenergies $E$ between $-\Delta$ and $\Delta$ as long as
$N_{\rm ac} a\approx N_{\rm zz} a\gg R \gg a$.  

\begin{figure}
\includegraphics[width=.95\columnwidth]{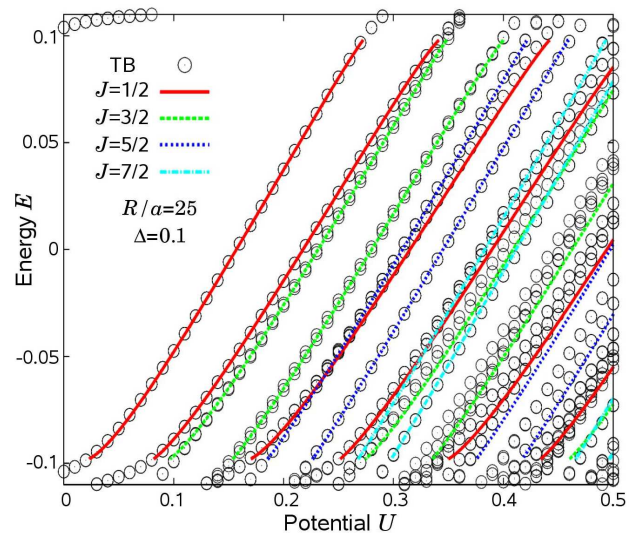}
\caption{(Color online) Comparison between the eigenenergies of the TB
  Hamiltonian (circles) of the graphene lattice and the solutions of
  the Dirac equation (lines) as a function of the potential $U$. The
  parameters used are $\Delta=0.1$ and $R=25\, a$.
\label{BoundStates2}} 
\end{figure}

Figure~\ref{BoundStates2} exhibits the eigenvalues of the TB
Hamiltonian~(\ref{TBham}) vs. the potential $U$ in the energy range
$|E| < \Delta$. For comparison, we show also the eigenenergy solutions
of the Dirac equation for the bound states as given by \eq{BS}. We
restrict the angular momentum to $J \le 7/2$ for the sake of clarity. 
There is an excellent agreement between the two approaches for small
angular momenta and small $U$. The angular momentum is not a good
quantum number in the 
lattice model, nevertheless, we can identify the energies in both
models for small angular momenta. If larger QD radii are considered,
the good agreement between the TB and the Dirac equation becomes better
even for higher values of $J$ (results not shown here because there
are too many eigenvalues in the figure, already for $R\sim 60\,a$). 
The reason for the slight deviations for small QD radii are the
irregularities at the QD border appearing in the lattice model but
which are not present in the continuum model. The corresponding
lattice QD is not a perfect circle but has a chiseled edge.  

\section{Transport properties of a graphene QD \label{SET}}
\subsection{Continuum model}
\begin{figure}
\includegraphics[width=8.5cm]{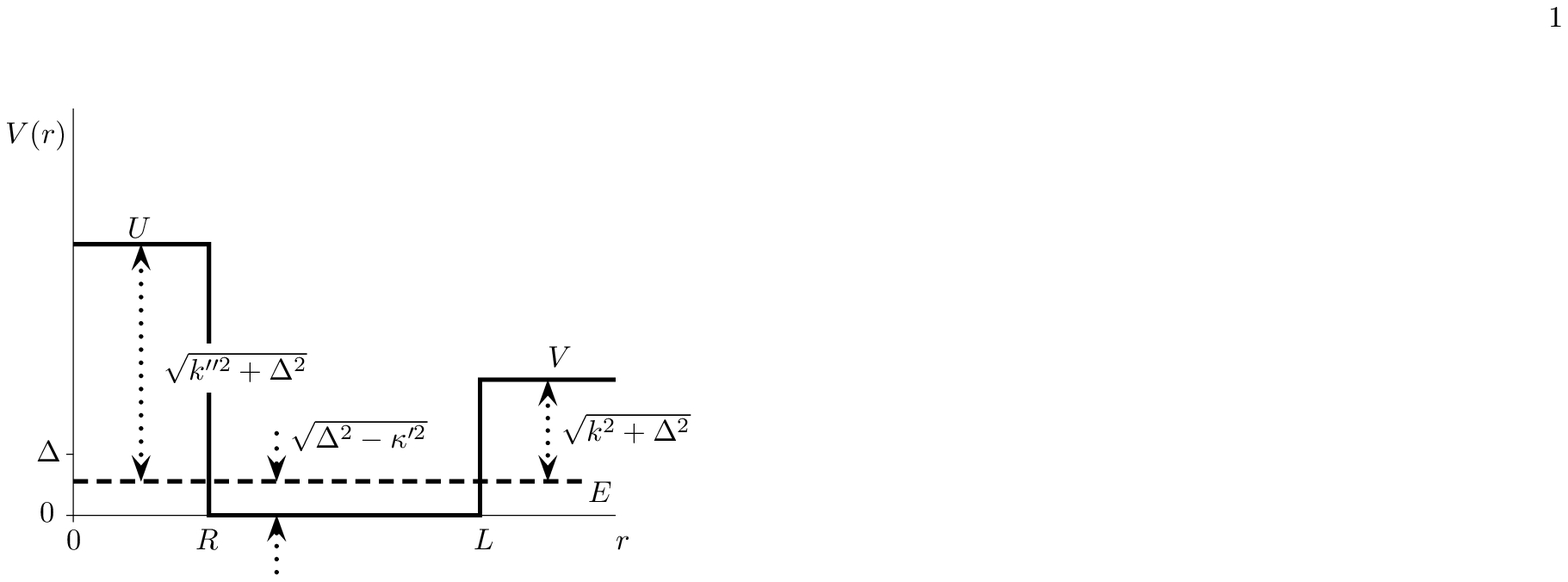}
\caption{(Color online) Potential landscape of a QD with radius $R$
  and an additional environment for $r>L$. The vertical dotted arrows
  indicate $|E-V(r)|$ in each of the three regions. For $R<r<L$, the 
  wavefunctions decay. Also, $k''$, $k'=i\kappa'$, and $k$ are the
  wavevectors inside the QD, outside, and in the environment region,
  respectively.} 
\label{QDE}
\end{figure}

In this section, we want to calculate the transport properties through
the circular QD. For this purpose, we need to introduce an environment
to the isolated QD so that the exponentially decaying bound states
are still finite when reaching the outer region. We take a radially
symmetric model with the following potential 
\be
V(r)=\left\{ \begin{array}{ll} U, & 0<r<R \\ 0, & R<r<L \\ V, &
  L<r \end{array} \right. . 
\ee
The profile of the potential, which now consists of three regions, 
is schematically shown in Fig.~\ref{QDE}. The radially symmetric
choice makes the analytic determination of the S-matrix possible.
Next, we explain our setup and sketch the derivation of the S-matrix. 
 
Compared to the isolated QD, there is a third region for the radial
component, $r>L$. Here, the potential $V$ is chosen such that the
wavefunctions are oscillatory. Then, we can have incoming and outgoing
waves in the environment and define a scattering cross-section. 
In the intermediate region, $R<r<L$, the wave functions decay, and
inside the QD, $0<r<R$, the energy can belong to bound states. This
corresponds to the regime II of the isolated QD.

Next, we turn to the definition of the scattering matrix 
$\mS$.\cite{Nov07a,KN07,Gui08}.  
For $r\to \infty$, we write the wave function as
\bea
\xi_{\pm}(\vx) &=& \frac{1}{\sqrt{2}} 
\sqrt{E-V\pm\Delta} \; e^{ikx} +  \NL
 & & \frac{1}{\sqrt{2\pi kr}} 
\sqrt{E-V\pm\Delta}\;f_{\pm}(\varphi) \;  e^{ikr}. 
\label{asy}
\eea
The first term is the incoming plane wave in $x$ direction, the second
the outgoing spherical wave. Furthermore, $f_{\pm}(\varphi)$ is the
scattering amplitude, with $f_{+}(\varphi) = e^{i\varphi}
f_{-}(\varphi)$, and $\varphi$ is the scattering angle. The two
components of $\xi$ individually solve the Dirac equation for graphene
and are normalized such that the incoming (plane wave) particle
current density is $k$, while the number of particles leaving per unit
time radially in the direction $\varphi$ is $d\varphi \frac{1}{2\pi} 
\left( f_{+}^*(\varphi) e^{i\varphi} f_{-}(\varphi) 
 + f_{-}^*(\varphi) e^{-i\varphi} f_{+}(\varphi) \right) $.
Thus, the differential scattering cross section is 
\be
\frac{d\sigma(\varphi)}{d \varphi} = \frac{1}{2\pi k}
\left( f_{+}^*(\varphi) e^{i\varphi} f_{-}(\varphi)
 + f_{-}^*(\varphi) e^{-i\varphi} f_{+}(\varphi) \right).
\ee
We then expand in angular momenta with coefficients $f_J$
\be
 f_{\pm}(\varphi) = \frac{1}{\sqrt{2}} 
              \sum_J e^{i\varphi (J\pm\frac{1}{2})-i\frac{\pi}{4}} \; f_J
\ee
and define the scattering matrix $\mS_J$ as $\mS_J = 1 + f_J$ to obtain
\be       \label{dwqs}
\frac{d\sigma(\varphi)}{d \varphi} = \frac{1}{2\pi k} \sum_{J,J'}
\; e^{i\varphi (J-J')} \; ({\mS}_{J'}^*-1) (\mS_{J}-1)  \;.
\ee

Next we sketch the calculation of $\mS_J$. In our geometry, the
general solution of the Dirac equation for fixed angular momentum
$J$ is given by 
\begin{eqnarray*}
\lefteqn{\xi_{\pm}(r) = }\\
& & \left\{ \begin{array}{ll}
{\mA}_{\pm} \; H^{(1)}_{J\pm\frac{1}{2}}(k r) 
+{\mB}_{\pm} \; H^{(2)}_{J\pm\frac{1}{2}}(k r), & L<r \\
{\mA}'_{\pm} \; K_{J\pm\frac{1}{2}}(\kappa' r) 
+{\mB}'_{\pm} \; I_{J\pm\frac{1}{2}}(\kappa' r), & R<r<L \\
{\mA}''_{\pm} \; J_{J\pm\frac{1}{2}}(k'' r), & r<R.
 \end{array} \right. 
\end{eqnarray*}
Here, the wave vectors are $k=\sqrt{(E-V)^2-\Delta^2}$, 
$\kappa'=\sqrt{\Delta^2-E^2}$, and $k''=\sqrt{(E-U)^2-\Delta^2}$.
Using in (\ref{asy})
\be 
 e^{ikr\cos(\varphi)} = \sum_{m=-\infty}^\infty 
   e^{i(\varphi+\frac{1}{2} \pi) m} \; J_m(kr) \;,
\ee
we can easily find the expression for ${\mA}_{\pm}$ and ${\mB}_{\pm}$ 
from the ansatz \eq{asy},
i.e., the relation between ${\mA}_{\pm}$ and ${\mB}_{\pm}$ and $\mS_J$.
Then, ${\mA}_{\pm}$ and ${\mB}_{\pm}$ are related to 
${\mA}'_{\pm}$ and ${\mB}'_{\pm}$, and these in turn to ${\mA}''_{\pm}$ 
by the matching conditions. 
Thus, the elements of the scattering matrix $\mS_J$ are determined. 
We skip the tedious but straight-forward (linear) algebra and just
quote the result 
\be
 \mS_J = - \frac{\det \mD^{(2)}}{\det \mD^{(1)}} \label{S}
\ee
and
{\setlength{\arraycolsep}{0pt}
\bea
&&\mD^{(1,2)} = \NL
 && \mm{cccc}{
 0&w'_+ K_{J_+}(x'_R) &w'_+ I_{J_+}(x'_R) & w''_+ J_{J_+}(x'')\\
 0&w'_- K_{J_-}(x'_R) &-w'_- I_{J_-}(x'_R) &w'' _- J_{J_-}(x'')\\
 w_+ H^{(2,1)}_{J_+}(x) &w'_+ K_{J_+}(x'_L) &w'_+ I_{J_+}(x'_L)&0\\
 w_- H^{(2,1)}_{J_-}(x) &w'_- K_{J_-}(x'_L)&-w'_- I_{J_-}(x'_L)&0} \NL 
\label{D12}
\eea
}
The appearance of the determinants is easily explained: they result
from the solution of the $4$ linear matching conditions ($2$ for
$\xi_{\pm}$ at $r=L$ and $2$ more at $r=R$). Again, $H^{(1,2)}_{m}(x)$,
$K_{m}(x)$, $I_{m}(x)$, and $J_{m}(x)$ are the standard Bessel
functions and we use the following abbreviations:
\bea
 &&w_\pm := \sqrt{|E-V\pm\Delta|} \;,\quad           \label{Abr}
 w''_\pm := \sqrt{|E-U\pm\Delta|} \;, \quad  \NL
 && w'_\pm := \sqrt{|E\pm\Delta|}\;, \quad
 J_\pm := J \pm \frac{1}{2}\;, \NL
 &&x := kL, \quad x'_{R; L} := \kappa' R;\ \kappa' L\;, 
\quad x'':= k'' R\;.
\eea

\begin{figure}[t]
\includegraphics[width=.95\columnwidth]{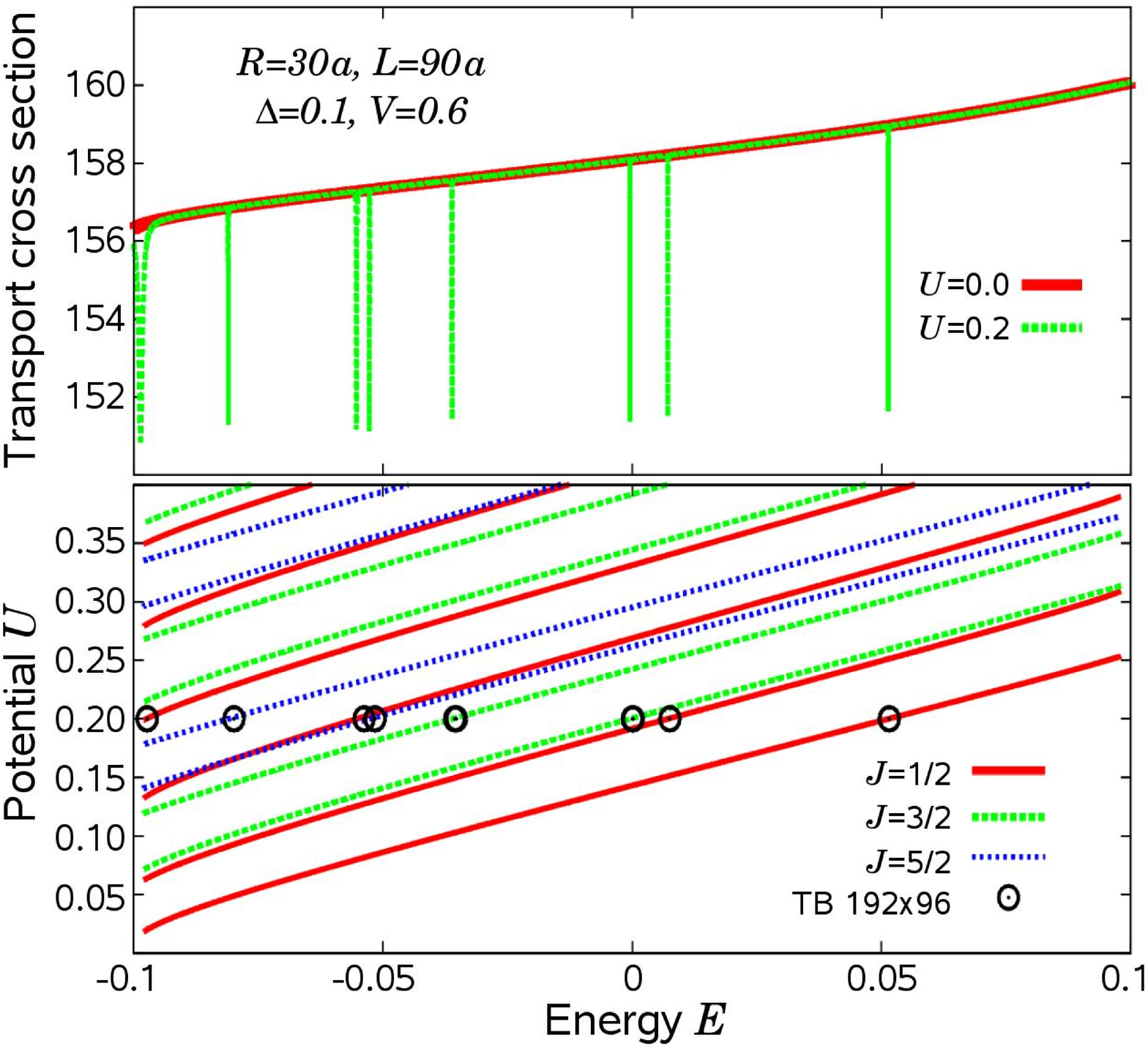}
\caption{(Color online) Transport cross-section of a graphene quantum
  dot ($R/a=30$, $\Delta=\pm 0.1$) with environment (upper panel). 
  The position of the peaks correspond to the energies of the bound
  states of an isolated graphene QD with dot potential $U=0.2$. They
  also agree with the results from the corresponding TB model (circles
  in the lower panel). The transport cross section (green dotted line)
  is shown in arbitrary units with the 'background' value (red solid
  line) given by the empty dot $U=0$. 
\label{Transport1}}   
\end{figure}

To make connection with the two-terminal conductance that can be
measured in experiments, we calculate the transport cross section
as\cite{KGG09}
\be
\sigma_{tr} := \int_{-\pi}^\pi d \varphi \; (1-\cos \varphi) \; 
 \frac{d\sigma(\varphi)}{d \varphi} .
\ee
For the scattering cross section from \eq{dwqs}  we use the $\mS_J$
computed according to Eqs.~(\ref{S}-\ref{Abr}). The results are shown
in Fig.~\ref{Transport1}, where 
we compare the bound state energies of an isolated graphene QD
of fixed size $R$ and for fixed potential $U$ with the positions of
the resonances in the transport cross section of the QD plus
environment. The bound state energies and the positions of the
resonances exactly coincide. When the energy of an incoming particle
matches that of a  bound state, the particle tunnels through the
QD and the cross section is small. The 'background' cross-section is
given by $\sigma_{tr}(U=0)$, i.e., by the cross section of an 'empty'
QD. 

\begin{figure}
\includegraphics[width=.95\columnwidth]{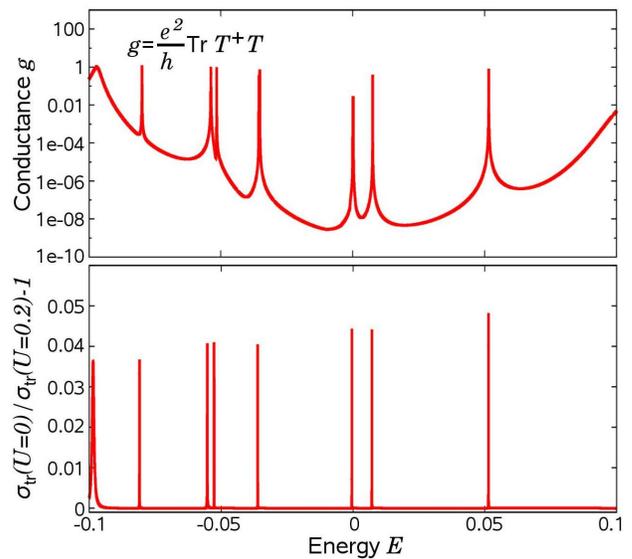}
\caption{(Color online) The two-terminal conductance $g$ (upper panel)
  calculated within a TB model compared to the normalized transport
  cross section (lower panel) of a continuum Dirac model. 
\label{Transport2}} 
\end{figure}

\subsection{Lattice model}
In this subsection, the two-terminal conductance $g$ through a
graphene quantum dot is calculated numerically within a lattice model
and compared with the results of the transport cross-section of the
continuum model. The two-terminal conductance\cite{ES81} is obtained
from the transmission matrix $T$ employing the transfer-matrix
method\cite{PMR92} according to    
\begin{equation}
g=\frac{e^2}{h} {\text{Tr}}\,T^{\dag}T=\frac{e^2}{h} \sum_{i}^{N}
(\cosh^2(\varepsilon_i/2))^{-1},
\end{equation}
where the $\varepsilon_i$ parameterize the eigenvalues of $T$, and $N$
is the number of open channels.
The electron transmission through a sample of width $L_y=96\,a$
and length $L_x=192\,a$ with two semi-infinite leads attached at $x=1$
and $x=L_x$ is calculated numerically. Periodic boundary conditions
are applied in the $y$ direction. The sublattice anisotropy, the dot
potential and radius are $\Delta/t=\pm 0.1$, $U/t=0.2$, and $R/a=30$,
respectively. The outcome is shown in 
Fig.~\ref{Transport2}, where the two-terminal conductance (upper panel)
and the relative transport cross-section (lower panel) are plotted 
as a function of energy. The latter is normalized according to  
$\sigma_{tr}(U=0)/\sigma_{tr}(U)-1$ while the former is displayed on a
log-scale in order to make the comparison easier. The reason is that
in contrast to the transport cross-section of the infinite system
considered in the continuum model, the two-terminal conductance, due
to the finite sample length in the lattice model, does not drop to
zero between the resonances that are associated with the bound
states. For the energetic positions of the resonances, an excellent
agreement is obtained between the two approaches. However, the
agreement between the two is good only as long as $R\gg a$. When the
QD radius is reduced, the positions of the transport peaks, calculated
within the continuum model, differ from those of the lattice model due
to the irregularities at the QD edges appearing in the latter approach.

It is clear that the electrostatic potential confining the QD cannot
be infinitely steep in real samples. Therefore, we also studied the
effect of linearly sloping dot potential steps. We find that the
electronic bound states of the QD shift linearly to smaller energies
when the slope of the QD walls decreases. Therefore, it can happen
that the bound states' energies fall below the lower edge of the gap,
$E=-\Delta$. Hence, in experiments the dot potentials should be
fabricated as steeply as possible.    

\begin{figure}
\includegraphics[width=8.0cm]{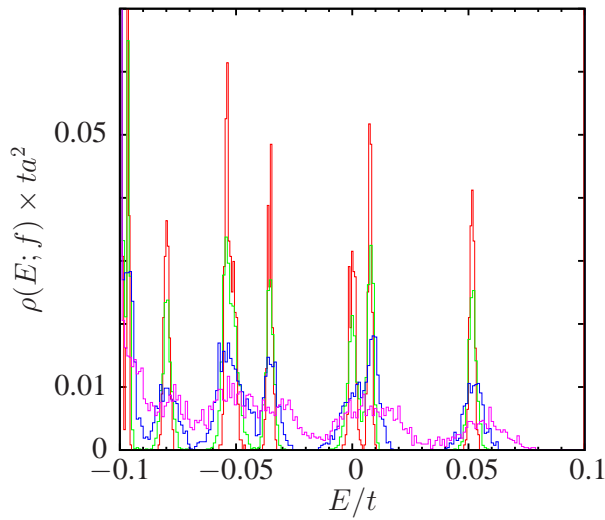}
\caption{(Color online) Influence of random magnetic flux disorder 
on the density of states of a circular QD with radius 30\,$a$ and dot
potential $U/t=0.2$. The sublattice difference is $\Delta/t=\pm 0.1$
as before. With increasing disorder, $f/(h/e)= \mathrm{0.005}$
(red), 0.01 (green), 0.02 (blue), and 0.05 (magenta), the peaks get
broadened and start to overlap.} 
\label{rflx}
\end{figure}

\section{\label{disorder}Influence of disorder}
Real graphene samples are usually subject to various disturbances and
imperfections. It is known that the electronic properties essentially
depend on the particular type of disorder present. In the following,
we study the influence of possible modifications that may affect the
bound states of an electrostatically confined graphene quantum dot.  
The lattice model Hamiltonian of the unperturbed system  (\ref{TBham})
is now modified to allow for diagonal and non-diagonal disorder. The
latter arises in graphene when either ripples or random magnetic
fields are present.\cite{Mea06,MG06,Mea07,GKV08,SM08a} The impact
of random on-site disorder that can be due to adsorbed atoms and
molecules attached to the bulk or to the dangling bonds along the 
edges of the graphene sheet, is described by diagonal
random disorder potentials   
\begin{equation}  
{\cal H}=\sum_{\vvr} (U_{\vvr}+\Delta_{\vvr}+\epsilon_{\vvr})\, 
c_{\vvr}^{\dagger}c_{\vvr}^{}-
t\sum_{\langle \vvr\ne\vvr'\rangle} e^{-i\theta_{\vvr\vvr'}}
c_{\vvr}^{\dagger}c_{\vvr'}^{}.
\end{equation}
Again, $t=2.7$\,eV is the nearest-neighbor transfer energy for
graphene, $\vvr$ denotes the sites of the carbon atoms, and $\langle
\vvr\ne\vvr'\rangle$ are pairs of those neighboring sites on a
bricklayer lattice\cite{SM08a} that are connected by bonds. The sample
width and the length are $L_y=96\,a$ and $L_x=192\,a$, respectively. 
Periodic boundary conditions (BC) are applied in the 
$x$ direction (length) and Dirichlet BC in the $y$ direction (width).
As before, $U_{\vvr}$ and $\Delta_{\vvr}$ define the quantum dot
potential and the sub-lattice anisotropy, respectively. The
on-site potentials $\epsilon_{\vvr}$ describe either the uncorrelated 
random bulk disorder or the one-dimensional edge-disorder. A
box-probability distribution $P(\epsilon_{\vvr}) = 1/W$ is assumed for
the random potentials $-W/2\le \epsilon_{\vvr} \le W/2$, where $W$ is
a measure of the disorder strength. The case of ripples and
corrugations is modeled by an equivalent random-magnetic-flux (RMF)
disorder model leading to complex phase factors that are defined by 
$\sum_{}\theta_{\vvr\vvr'}=2\pi e\phi_{\vvr}/h$, where the sum is
taken along the bonds around a given plaquette situated at $\vvr$ with
the random flux $\phi_{\vvr}$ piercing through it. Without a QD, the
chiral symmetry is conserved\cite{OGM06,MS07,SM08a,Sch09} in this
disorder model. 
The random fluxes are also drawn at random from a box distribution
$-f/2\le\phi_{\vvr}\le f/2$ with zero mean, and $h/e$ is the magnetic
flux quantum. The RMF-disorder strength $f$ varies between $0\le
f/(h/e)\le 1$. 

\begin{figure}
\includegraphics[width=8.0cm]{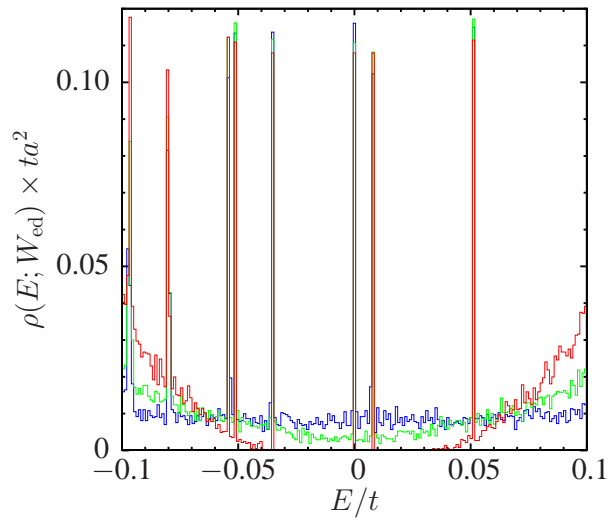}
\caption{(Color online) The density of states $\rho(E;W_{\rm ed})$
  within the gap arising from the sublattice difference $\Delta/t=\pm
  0.1$ in the presence of edge disorder of strengths $W_{\rm ed}/t= 
  \mathrm{0.1}$ (red), 0.2 (green), and 0.5 (blue). The bound states
  energies of the quantum dot (radius 30\,$a$, potential strength
  0.2\,$t$) remain almost unaffected by the edge disorder.}
\label{edgdis}
\end{figure}

The result of random-flux disorder is shown
in Fig.~\ref{rflx}, where the density of states (DOS) of a graphene
sheet with quantum dot and sublattice anisotropy is plotted for
energies around the Dirac point. The sublattice difference is taken to
be $\Delta/t=\pm 0.1$, the quantum dot potential is $U/t=0.2$, and the
QD radius is $R/a=30$. The peaks of the bound states are still
clearly visible in the case of small disorder strength
$f/(h/e)=0.005$. With increasing disorder, however, the peaks get
broadened and disappear for $f/(h/e)=0.05$. Therefore, the sharp bound
energy states of a clean graphene QD can be destroyed in the
presence of sufficiently strong ripple disorder.  
 
A completely different behavior is obtained in the case of one-dimensional
edge-disorder only.\cite{EZHH08,MCL09,KOQF10} 
This can be seen in Fig.~\ref{edgdis} where again
the DOS is shown for energies within the gap $\Delta/t=\pm 0.1$ of a 
quantum dot with potential $U/t=0.2$. With increasing edge-disorder
strength from $W_{\rm ed}/t=0.0$ to $W_{\rm ed}/t=0.5$, only the DOS
of the peaks lying outside the gap are broadened and eventually
completely fill the energy regions between the sharp bound states that
remain unaffected.   
Hence, electrostatically confined graphene QD do not deteriorate in the
presence of edge-disorder and remain suited for experimental
spectroscopic studies. This behavior is due to the exponential decay
of the edge states which do not get mixed with the localized bound
states of the QD even in the presence of additional edge-disorder 
potentials.

\begin{figure}
\includegraphics[width=8.0cm]{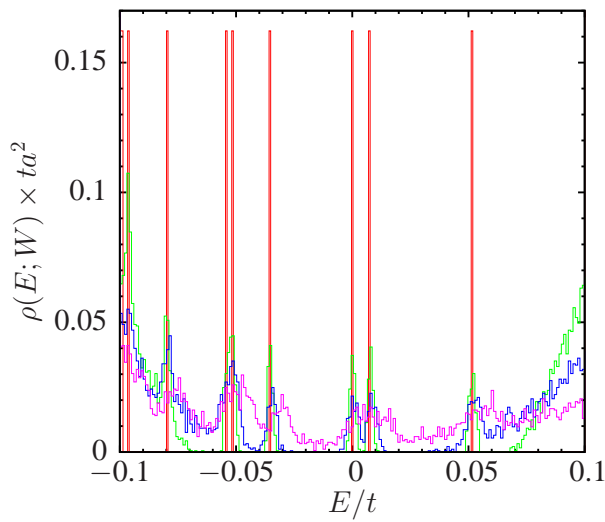}
\caption{(Color online) The density of states $\rho(E;W)$ showing the
  bound states of the quantum dot (radius 30\,$a$, potential strength
  0.2\,$t$) within the gap arising from the sublattice difference
  $\Delta/t=\pm 0.1$ in the presence of on-site disorder of strengths
  $W/t= \mathrm{0.0}$ (red), 0.05 (green), 0.1 (blue), and 0.2
  (magenta). The sharp DOS-peaks broaden and finally disappear
  with increasing disorder strength.}
\label{nondiagdis}
\end{figure}

When in addition to edge disorder the random on-site potentials are also
allowed to affect the bulk sites, the DOS of the quantum dot's bound
states get broadened with increasing disorder strength as can be seen
in Fig.~\ref{nondiagdis}. As in the case of the random-flux disorder,
but in stark contrast to the edge-disorder only situation, the sharp
peaks broaden and finally disappear for disorder strength $W/t>0.2$. 
A similar behavior is seen in the two-terminal conductance where, with
increasing disorder strength, the single sharp resonances are replaced
by a cluster of resonances which essentially depend on the particular 
disorder realization. So, uncorrelated short-range random potentials
of sufficient strength destroy the discreteness of the QD energies. In  
order to fabricate electrostatically defined graphene quantum dots,
only samples that are clean, at least in the bulk area, can be used,
whereas disorder at the edges is of minor importance.

\section{Conclusions}
The electronic properties of circular graphene quantum dots, which
can be created electrostatically in the presence of a sublattice
asymmetry, were investigated with the aid of two different models. 
First, we used a continuum model described by a Dirac-type equation
that can be solved analytically, and second, we applied a
tight-binding lattice model, which was evaluated numerically. 
The dependence of the electronic bound states
and the transport cross-section on the electrostatic potential and
size of the quantum dot were calculated. Both the spectra of bound
quantum dot states and the electric transport through the QD show a
very good agreement between the two models when the radius of the QD
is much larger than the carbon-carbon distance $a$. For smaller dot
radii, the agreement is always reasonable for the low lying bound
states, but gets worse at higher energies where in the continuum model  
the large angular momenta essentially contribute. In experiments, QDs
of sizes 20--50\,nm are fabricated\cite{Pea08,TCAG09,Gea09} which means
$R/a\approx 70-176$. For such radii, the agreement between the two
models is very good even for bound states with high values of the
angular momentum $J$.   

The results of the transport cross-section for a QD with attached
environment calculated within the infinite continuum model coincided
with the numerically evaluated two-terminal conductance obtained for a
finite graphene sample with attached semi-infinite leads described by
a tight-binding lattice model.  
Within the TB model, we also studied the influence of a sloping dot
potential, which caused the energy levels to shift to lower energies,
and of bulk and edge disorder on the bound states of electrostatically
defined graphene quantum dots. 

The presence of disorder severely influences the quantum dot's
spectral and transport properties. Only in the case of one-dimensional
uncorrelated random edge-disorder, do the peaks of the density of states
corresponding to the bound states remain sharp. For on-site or
random-magnetic-flux (ripple) bulk disorder, the peaks of the bound
states in the DOS are broadened and finally disappear with
increasing disorder strength, and thus the quantum dot loses its 
experimentally observable characteristic spectral fingerprint.


\end{document}